\documentstyle[aps,twocolumn,psfig]{revtex}
\begin{document}
\draft
\title{NMR Quantum Automata in Doped Crystals}
\author{Haiqing Wei, Xin Xue and S. D. Morgera}
\address{
Department of Electrical and Computer Engineering, McGill
University, Montreal, Quebec, Canada H3A 2A7 \\
E-mail: dhw@insl.mcgill.ca}
\maketitle
\begin{abstract}
In a lattice ${\cal L}$ of nuclear spins with ABCABCABC$\cdots$ type 
periodic structure embedded in a single-crystal solid, each ABC-unit can
be used to store quantum information and the information can be moved 
around via some cellular shifting mechanism. Impurity doping marks a 
special site D$\not\in{\cal L}$ which together with the local spin lattice
constitute a quantum automaton where the D site serves as the input/output
port and universal quantum logic is done through two-body interactions
between two spins at D and a nearby site. The novel NMR quantum computer
can be easily scaled up and may work at low temperature to overcome the
problem of exponential decay in signal-to-noise ratio in room temperature 
NMR.
\end{abstract}

In the request of physical hardwares for quantum computation, nuclear spin
systems hold strong promise due to their inherent ability of coherence
preserving [1]. A nuclear spin computer would be possible if each spin 
could be individually addressed, but the technical implementation with the 
nanoprobe method for example, is rather difficult. A recent breakthrough 
in bulk quantum computation using solution nuclear magnetic resonance 
(NMR) has brought quantum computers into laboratory [2-4]. NMR computers  
in solution are however limited to small scale for some reasons. Firstly,
to scale up one has to find or design a larger molecule and completely
explore interactions among its nuclear spins as well as chemical shifts. 
That is a daunting project unless the molecule has a simple structure as
in the case of periodic polymer, even so the polymer may not stay at 
simple structure in solution, it may fold. Secondly, the signal-to-noise
ratio degrades exponentially in room temperature NMR. Simply cooling the
solution down won't be good as it may end up as some amorphous solid so
the powder-broadening effect in solid-state NMR [5] fails quantum
manipulation of nuclear spins, unless, one can make the solution 
solidified into a single crystal which is rather unlikely. Then why not
start every thing from a single crystal?

Just for simplicity in this letter, all nuclei, if spinning, are assumed
to be spin-1/2 ones. Consider a lattice ${\cal L}$ of nuclear spins having
periodic structure  of ABCABCABC$\cdots$ type in three dimensions, where
A, B and C are three distinguished sites of nuclear spin. The nuclei at
three sites are not necessarily different, they may be all protons $^1$H
at distinguished chemical environments for example. ${\cal L}$ may be
embedded in a crystal lattice of some solid compound, there should be
other lattice points $\not\in{\cal L}$ with non-spinning nuclei (or their
spins are irrelevant) sitting on site. A two-dimensional ABC-periodic spin
lattice is shown in Fig.1. Each ABC-unit can be used to store quantum
information by setting one of the spin up or down (even in quantum
superposition of the two) and the information can be moved around via some
cellular shifting mechanism [6]. To be specific, if some information is
originally written on an A site of the one-dimensional chain
ABCABCABC$\cdots$, exploiting two-body coupling between adjacent spins
enables the quantum swap operation and cascading swaps of
A$\rightleftharpoons$B, B$\rightleftharpoons$C, C$\rightleftharpoons$A,
A$\rightleftharpoons$B$\cdots$ keeps shifting the information to the right
and can post it onto any desired site [6]. Indeed, the quantum swap
operation can be achieved by cascading three quantum controlled-NOT gates
[7]. If the crystal is doped such that the donor nucleus D with spin 1/2 
replaces some non-spinning nucleus in the proximity of an A site of
${\cal L}$ as shown in Fig.1, then the special impurity site can serve as
the input/output port, and a small piece of sublattice ${\cal
L}'\subset{\cal L}$ in the local region, about tens of ABC-units along all
three dimensions, free of any defect and other impurity, provides a large
quantum memory with a number of qubits over thousands. Therefore impurity
doping may induce large-scale quantum automata in a single crystal and the
whole crystal contains a huge ensemble of identical NMR quantum computers
which share the same chemical anisotropy so to get rid of the
powder-broadening effect and perform perfectly at low temperature. The
next question is how to bring them into operation.

\begin{figure}
\begin{center}
\ \psfig{file=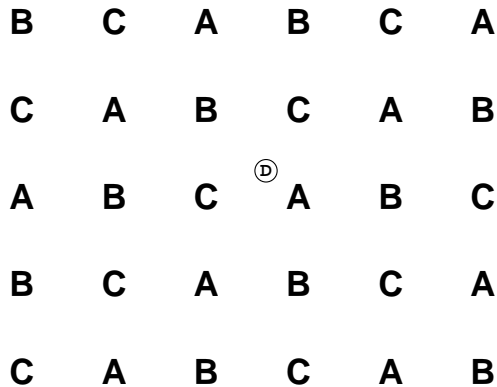,height=5cm,width=6.4cm}
\vspace{3mm}
\caption{A two-dimensional nuclear spin lattice with
the ABCABCABC$\cdots$ type periodicity.} 
\label{fig1}
\end{center}
\end{figure}

If the crystal is sufficiently cooled down and subject to strong magnetic
field for some while longer than the spin-relaxation time, maybe hours
even days but tolerable, most nuclear spins will relax to the lower level
which is the logic $|0\rangle$ state. A quantum automaton of size $N^3$,
that is a donor nucleus D together with the local piece of sublattice
${\cal L}'\subset{\cal L}$ containing $N$ ABC-units along all three
dimensions, has a large probability to be perfectly initialized with all
spins at the $|0\rangle$ state. Even at higher temperature that 
relaxation is less efficient in spin-polarization, a recent proposal [8]
shows how quantum manipulation can efficiently and perfectly initialize 
a subset of the quantum register. The perfectly initialized machines are
those that actually work in ensemble computation. Of course most of the
sublattice ${\cal L}'$ should be free of any defect and impurity other
than D. With careful material purification and crystal growth, it is not
difficult to make $N$ on the order of tens and more. For simplicity one
may assume that the spin lattice possess' an orthorhombic symmetry that
the three ABC-periodic chains along the $x$, $y$ and $z$ directions
respectively are associated with distinguished resonant frequencies to
drive the cellular shifting operation. Starting from the perfectly 
initialized automaton, information can be input by setting the D-spin to
the desired state then transferring the information to the nearest A-spin
via the quantum swap operation, and consequent cellular shifting
operations along the ABC-chains post the information to any prescribed
site (in 3-dimension). All these can be done selectively if the specified
quantum transitions are driven by external fields with distinguished
frequencies. One problem not to be overlooked is that the D-impurity may
affect some nearby A,B,C-spins so that the resonant pulses to drive the
quantum swaps among them are detuned (in frequency) from the standard
pulses that drive the same operations among these far-off spins. This 
problem however is easily solved by applying two (maybe more) sets of
pulses, one standard set to drive the far-off spins and the other set to
drive the near-D spins, again the desired selectivity is due to the
difference in resonant frequencies. After the required information is
loaded and the D-spin is reset to the $|0\rangle$ state, the computer is
ready for further quantum manipulation. Actually many quantum algorithms
start from the $|00\cdots 0\rangle$ state and do not need to load
information in the beginning. One major achievement in quantum computing
ensures that one-bit rotation and two-bit conditional gate are enough for
universal quantum computation [9-11], therefore one needs only figure out
how to  implement one-bit and two-bit gates on the present quantum
automaton. Any quantum operation on a single prescribed qubit X can be
done by first moving X via cellular shifting to the A-site which is the
nearest to the D-impurity, then applying a proper pulse to drive the
quantum operation on this A-spin which has a unique resonant frequency due
to its proximity to D, finally it may be preferred to move all qubits back
to their original sites. Any conditional quantum logic on two specified
qubits X and Y can be done by first moving X to the A-site nearest to D
and transferring it to the D-site via quantum swap A$\rightleftharpoons$D,
then bringing Y to the A-site nearest to D and doing the conditional logic
between D and this A, and then moving the Y qubit back to its original
site thus resetting all qubits back to position except that the X qubit
and the state of the D-spin are interchanged. The final restoration of
position is achieved by again moving the X qubit to the A-site nearest to
D and swapping A$\rightleftharpoons$D then posting the X qubit back to
site. Upon completion of computation, the state of any qubits can be 
measured by moving it to the A-site nearest to D, then swapping 
A$\rightleftharpoons$D, finally measuring the D-spin. One concludes that
the proposed quantum automata in crystal support all necessary ingredients
for quantum computation thus qualify as universal quantum computers.

In summary, the possibility and advantage of building NMR quantum automata
in doped crystals are discussed. Both the material preparation and the 
solid-state NMR are well within the reach of state-of-the-art technology,
the proposed quantum automata are especially suitable for low temperature
operation so to overcome the problem of exponential decay in
signal-to-noise ratio in room temperature NMR. Even at higher temperature,
the ensemble of quantum automata in a single crystal should be an 
excellent model for bulk quantum computation and may serve as a good
candidate to test the recent proposal of scalable NMR quantum computation
using some spin-polarization techniques [8], such as optical pumping [12]
and electron spin-flip scattering [13,14]. This letter is intended to call
for experimental studies on solid-state NMR with intentionally doped
crystals so to bring our dream of powerful quantum computers into reality
(or break it at all?).

\end{document}